\documentclass[twocolumn]{aastex63}
\usepackage{xcolor}
\usepackage{makecell}
\usepackage{pbox}

\received{\today}
\submitjournal{ApJ}
\shorttitle{ASASSN-14ko}
\shortauthors{Payne et al.}
\graphicspath{{./}{figures/}}

\def\gal{ESO 253$-$G003}

\begin{document}

\title{The Rapid X-ray and UV Evolution of ASASSN-14ko}

\correspondingauthor{Anna V. Payne}
\email{avpayne@hawaii.edu}

\author[0000-0003-3490-3243]{Anna V. Payne}
\altaffiliation{NASA Graduate Fellow}
\affiliation{Institute for Astronomy, University of Hawai\`{}i at Manoa, 2680 Woodlawn Dr., Honolulu, HI 96822}

\author[0000-0003-4631-1149]{Benjamin J. Shappee}
\affiliation{Institute for Astronomy, University of Hawai\`{}i at Manoa, 2680 Woodlawn Dr., Honolulu, HI 96822}

\author[0000-0001-9668-2920]{Jason T. Hinkle}
\affiliation{Institute for Astronomy, University of Hawai\`{}i at Manoa, 2680 Woodlawn Dr., Honolulu, HI 96822}

\author[0000-0001-9206-3460]{Thomas~W.-S.~Holoien}
\altaffiliation{NHFP Einstein Fellow}
\affiliation{The Observatories of the Carnegie Institution for Science, 813 Santa Barbara St., Pasadena, CA 91101, USA}

\author[0000-0002-4449-9152]{Katie~Auchettl}
\affiliation{School of Physics, The University of Melbourne, Parkville, VIC 3010, Australia}
\affiliation{ARC Centre of Excellence for All Sky Astrophysics in 3 Dimensions (ASTRO 3D)}
\affiliation{Department of Astronomy and Astrophysics, University of California, Santa Cruz, CA 95064, USA}

\author[0000-0001-6017-2961]{Christopher~S.~Kochanek}
\affiliation{Department of Astronomy, The Ohio State University, 140 West 18th Avenue, Columbus, OH 43210, USA}
\affiliation{Center for Cosmology and AstroParticle Physics, The Ohio State University, 191 W.\ Woodruff Ave., Columbus, OH 43210, USA}

\author{K.~Z.~Stanek}
\affiliation{Department of Astronomy, The Ohio State University, 140 West 18th Avenue, Columbus, OH 43210, USA}
\affiliation{Center for Cosmology and AstroParticle Physics, The Ohio State University, 191 W.\ Woodruff Ave., Columbus, OH 43210, USA}

\author[0000-0003-2377-9574]{Todd~A.~Thompson}
\affiliation{Department of Astronomy, The Ohio State University, 140 West 18th Avenue, Columbus, OH 43210, USA}
\affiliation{Center for Cosmology and AstroParticle Physics, The Ohio State University, 191 W.\ Woodruff Ave., Columbus, OH 43210, USA}

\author[0000-0002-2471-8442]{Michael A. Tucker}
\altaffiliation{DOE CSGF Fellow}
\affiliation{Institute for Astronomy, University of Hawai\`{}i at Manoa, 2680 Woodlawn Dr., Honolulu, HI 96822}

\author{James D. Armstrong}
\affiliation{ Institute for Astronomy, University of Hawai\`{}i, 34 Ohia Ku St., Pukalani, HI 96768, USA}

\author[0000-0003-0442-4284]{Patricia T. Boyd}
\affiliation{Laboratory for Exoplanets and Stellar Astrophysics, NASA Goddard Space Flight Center, Greenbelt, MD 20771, USA}

\author{Joseph~Brimacombe}
\affiliation{Coral Towers Observatory, Cairns, QLD 4870, Australia}

\author{Robert~Cornect}
\affiliation{Moondyne Observatory, 61 Moondyne Rd, Mokine 6401 WA, Australia}

\author[0000-0003-1059-9603]{Mark E. Huber}
\affiliation{Institute for Astronomy, University of Hawai\`{}i at Manoa, 2680 Woodlawn Dr., Honolulu, HI 96822}

\author[0000-0001-8738-6011]{Saurabh W. Jha}
\affiliation{Department of Physics and Astronomy, Rutgers, the State University of New Jersey, 136 Frelinghuysen Road, Piscataway, NJ 08854, USA}

\author[0000-0002-7272-5129]{Chien-Cheng Lin}
\affiliation{Institute for Astronomy, University of Hawai\`{}i at Manoa, 2680 Woodlawn Dr., Honolulu, HI 96822}

\begin{abstract}
    ASASSN-14ko is a recently discovered periodically flaring transient at the center of the AGN \gal{} with a slowly decreasing period. Here we show that the flares originate from the northern, brighter nucleus in this dual-AGN, post-merger system. The light curves for the two flares that occurred in May 2020 and September 2020 are nearly identical over all wavelengths. For both events, \textit{Swift} observations showed that the UV and optical wavelengths brightened in unison. The effective temperature of the UV/optical emission rises and falls with the increase and subsequent decline in the luminosity. The X-ray flux, in contrast, first rapidly drops over $\sim$2.6 days, rises for $\sim$5.8 days, drops again over $\sim$4.3 days and then recovers. The X-ray spectral evolution of the two flares differ, however. During the May 2020 peak the spectrum softened with increases in the X-ray luminosity, while we observed the reverse for the September 2020 peak.   
    
\end{abstract}

\section{Introduction}

Active galactic nuclei (AGN) normally vary stochastically across the electromagnetic spectrum, and the variability can be reasonably well-modeled as a damped random walk (e.g., \citealt{kelly08}, \citealt{kozlowski10}, \citealt{macleod10}, \citealt{zu13}). Occasionally, AGN can undergo flares, or outbursts, in which their brightness increases dramatically over a finite period of time. These intense brightening events have been attributed to changes in inflow or accretion disk instabilities surrounding the central supermassive black hole (e.g., \citealt{kawaguchi1998}), or to tidal disruption events (TDEs), in which a star becomes ripped apart as it passes within the tidal radius of its host SMBH (\citealt{hills75}, \citealt{rees88}, \citealt{evans89}, \citealt{phinney89}).  These nuclear transients show diverse behavior and potentially new classes are still being discovered (e.g., \citealt{trakhtenbrot2019nature}, \citealt{frederick2020}, \citealt{neustadt2020}, \citealt{vanvelzen2021}).

In \citet{payne2020} we reported the discovery of ASASSN-14ko, which has shown flares with a mean period of $P_0 = 114.2\pm0.4$ days and a period derivative of $\dot{P} = -0.0017\pm0.0003$ \citep{payne2020}. These periodic flares have been observed since 2014 by the All-Sky Automated Survey for Supernovae (ASAS-SN, \citealt{shappee14}, \citealt{kochanek17}). No prior available data appear to constrain when these flares first began, so it is possible that they have been occurring undetected for decades, or longer. 

MUSE data presented in \citet{tucker2020} revealed that the host, \gal{}, is a  complex merger remnant involving two AGN and a larger tidal arm. Both nuclei are classified as AGN by several diagnostics, but the brighter northeastern nucleus exhibits asymmetric broad-line emission while the fainter southwestern nucleus shows narrow-line emission.   

As discussed in \citet{payne2020}, a likely explanation of the available data is that ASASSN-14ko is a repeating partial TDE as opposed to a SMBH binary or SMBH+perturbing star binary. \citet{sukova2021} also found some similarities between ASASSN-14ko's flares and the accretion rate simulations they found in GRMHD simulations of an orbiting star interacting with an accretion disk. Many aspects of the data, such as the similarities in the optical flares since 2014 and the modest mass loss rates, seem most consistent with the repeating partial TDE explanation, although interpreting each flare as one pericenter passage means each flare cannot be truly identical because precession would slowly change the orbital geometry, and over tens of encounters the total mass loss ceases to be modest. Making ASASSN-14ko a SMBH binary system seems unlikely due to the lack of emission line velocity shifts, but this could also be explained if the system is face-on or as a result of poor temporal sampling.

\citet{payne2020} analyzed data from the six years of the ASAS-SN survey and multi-wavelength data from the May 2020 flare. In this paper, we present analysis of the subsequent flare, which occurred as predicted in September 2020. This new flare presented the opportunity to compare the X-ray, UV, and optical emission between individual flares for the first time, and to further examine how the flares fit into the model as a repeating partial TDE. In Section \ref{observations}, we discuss the photometric and spectroscopic data used in this paper. In Section \ref{northern_nuclei}, we show that the northern, brighter nucleus in \gal{} is the origin of the flares. In Section \ref{xrayuvotevolution}, we analyze the September 2020 light curves and spectra, and we compare them to the previous flare in May 2020. Finally, in Sections \ref{discussion} and \ref{conclusions}, we discuss the two flares in the context of TDEs and summarize the results of this paper. Throughout this paper, we assume a flat $\Omega_m = 0.3$ universe, leading to a luminosity distance of $\approx 188~\rm{Mpc}$ and a projected scale of $\approx 0.85 ~\rm{kpc}/\rm{arcsec}$. We also assume a Galactic extinction of $\text{A}_{\text{V}} = 0.118$ mag \citep{schlafly11}.

\section{Observations} \label{observations}

Using our mean period $P_0 = 114.2 \pm 0.4$ days and period derivative $\dot{P} = -0.0017 \pm 0.0003$ from \citet{payne2020}, we predicted the subsequent flare would peak in the $g$-band on UT 2020-09-7.4 $\pm$ 1.1, and the optical peak time measured in Section \ref{uvopticalevol} was consistent with the prediction.  Anticipating that the multi-wavelength evolution would follow a similar trend as the May 2020 event, we scheduled X-ray, UV, and optical observations around the predicted optical peak.  

\subsection{ASAS-SN Photometry}

ASAS-SN is an ongoing all-sky survey to discover supernovae and other transient phenomena. The 20 robotic telescopes at five sites in both the northern and southern hemispheres are hosted by Las Cumbres Observatory Global Telescope (LCOGT, \citealp{brown13}). Each of the telescopes consists of four 14-cm aperture Nikon telephoto lenses with $8.\!\!''0$ pixels and a 4.5$^{\circ}$ $\times$ 4.5$^{\circ}$ field of view. All telescopes obtain data in the $g$-band. 

The data were reduced using a fully-automated pipeline based on the ISIS image subtraction package \citep{alard98, alard00}. Under most circumstances, a single photometric epoch represents three combined dithered 90-second image exposures with a 4.5$\times$4.5 degree field-of-view which are then subtracted from a reference image. We then used the IRAF package \texttt{apphot} to perform aperture photometry with a 2-pixel, or approximately $16.\!\!''0$, radius aperture on each subtracted image, resulting in a differential light curve. The AAVSO Photometric All-Sky Survey (APASS; \citealp{henden15}) was used to calibrate the photometry. All low-quality ASAS-SN images were inspected by eye to remove data affected by clouds or other systematic problems. 

\subsection{Swift XRT \& UVOT Photometry}

Using our prediction of the optical peak time and the expected multi-wavelength trend based on the May 2020 flare, we requested \textit{Swift} UltraViolet/Optical Telescope (UVOT, \citealp{roming05}) ToO observations (ToO ID: 14389, 14488, PI: Payne). The UVOT data were obtained in six  filters \citep{poole08}: $V$ (5468 \AA), $B$ (4392 \AA), $U$ (3465 \AA), $UVW1$  (2600 \AA), $UVM2$  (2246 \AA),  and $UVW2$ (1928 \AA). We used the HEAsoft (\hspace{-1mm}\citealt{heasarc2014}) software task \textit{uvotsource} to extract the source counts using a $16.\!\!''0$ radius aperture and used a sky region of $\sim$ $40.\!\!''0$ radius to estimate and subtract the sky background. This aperture size was used to match the ASAS-SN photometry. All fluxes were aperture corrected and converted into magnitudes and fluxes using the most recent UVOT calibration (\citealt{poole08}, \citealt{breeveld10}), and corrected for Galactic extinction. To properly isolate and measure the transient flux in each epoch, the quiescent host fluxes were measured in the same aperture and subtracted. We converted the \textit{Swift} UVOT $B$ and $V$ magnitudes to Johnson $B$ and $V$ magnitudes using the standard color corrections\footnote{\url{https://heasarc.gsfc.nasa.gov/docs/heasarc/caldb/swift/docs/uvot/uvot_ caldb_coltrans_02b.pdf}}.

\subsection{Las Cumbres Observatory Global Telescope Photometry \& Spectroscopy}

We obtained photometric observations from Las Cumbres Observatory Global Telescope (LCOGT, \citealt{brown13}) using the 1-m telescope at Siding Spring Observatory in New South Wales, Australia. The data consisted of $B$-, $V$-, $g'$-, and $r'$- observations. Aperture magnitudes were obtained using a $16.\!\!''0$ radius aperture using the IRAF \texttt{apphot} package using an annulus to estimate and subtract background counts. The data were calibrated using stars with APASS DR 10 magnitudes, and the aperture magnitudes were corrected for Galactic extinction. The quiescent host fluxes were measured in the same aperture and subtracted.

\subsection{Pan-STARRS Photometry} 
Located atop Haleakal$\mathrm{\bar{a}}$ on Maui, the Pan-STARRS1 telescope has a 1.8-m diameter primary mirror with a wide-field 1.4 gigapixel camera consisting of sixty Orthogonal Transfer Array devices, which has carried out a set of  synoptic imaging sky surveys with resulting catalog data products (\citealt{chambers16}, \citealt{flewelling2020}). Pan-STARRS1 data are obtained using $grizy_{P1}$ filters, similar to the SDSS filters \citep{abazajian09}. The Pan-STARRS1 photometric system is discussed in detail in \citet{tonry2012}.

The Image Processing Pipeline (IPP, see details in \citealt{magnier2020}) processes the Pan-STARRS1 data. The processing steps include device ``de-trending," a flux-conserving warping to a sky-based image plane, masking and artifact location involving bias and dark correction, flat-fielding, and illumination correction through rastering sources across the field of view \citep{waters2020}. There were 28 epochs of ASASSN-14ko Pan-STARRS1 data between 2014-2020. These data covered times both during quiescence and during separate flares that peaked in the optical on MJD $57651.9^{+3.0}_{-3.0}$, $57761.4^{+9.0}_{-3.8}$, and $58878.4^{+0.3}_{-0.6}$ \citep{payne2020}. We used the difference imaging technique created by IPP to obtain  stacked Pan-STARRS1 $3\pi$ data reference images using data taken in quiescence. Those reference images were then subtracted from the images taken during the prior outbursts to isolate the transient (e.g., \citealt{huber2015}).  

\subsection{SALT Spectroscopy}

We used the 10-m Southern African Large Telescope (SALT, \citealt{buckley2006}) with the Robert Stobie Spectrograph (RSS, \citealt{burgh2003}, \citealt{kobulnicky2003}) to obtain optical spectra during the flare. The data were obtained with the $1.\!\!''5$ slit on UT 2020-09-04 and UT 2020-09-16, which corresponded to $\sim$1 day prior and $\sim$11 days after to the measured optical $g-$band peak, respectively. The slit position was orientated with a position angle of $-1^{\circ}$ for the first spectrum and $30^{\circ}$ for the second spectrum. The data were then reduced using standard procedures, including bias subtraction, flat-fielding, wavelength calibration, sky subtraction, and flux calibration.

\subsection{Amateur Astronomer Photometry}

Data were collected using a 41-cm telescope at Savannah Skies Observatory from Queensland, Australia. The 180-second exposures were obtained using the $B$-, $V$-, and $R_C$- filters, and the data were bias and dark subtracted. Data were also taken at Moondyne Observatory located east of Perth, Australia, using a 0.4-m telescope with AOX adaptive optics. Guided $B_C$-, $V$-, $G_S$-, and $R_S$-band images were reduced and calibrated with bias and dark subtraction and flat-field normalization. Data were taken twice nightly over the duration of the event with 120 and 600 second exposures, and then aligned using background stars.

\section{The Brighter Nucleus is the Origin of the Flares} \label{northern_nuclei}

Using the spatially resolved MUSE IFU spectra, \citet{tucker2020} showed that the nuclear region of \gal{} houses two AGN separated by $1.\!\!''7$ as part of a post-merger system. While ASAS-SN's astrometry is generally accurate to $1.\!\!''0$, it is  not ideal for conclusively determining which nucleus is the origin of the outbursts. 

\begin{figure}
    \centering
    \includegraphics[width=\linewidth]{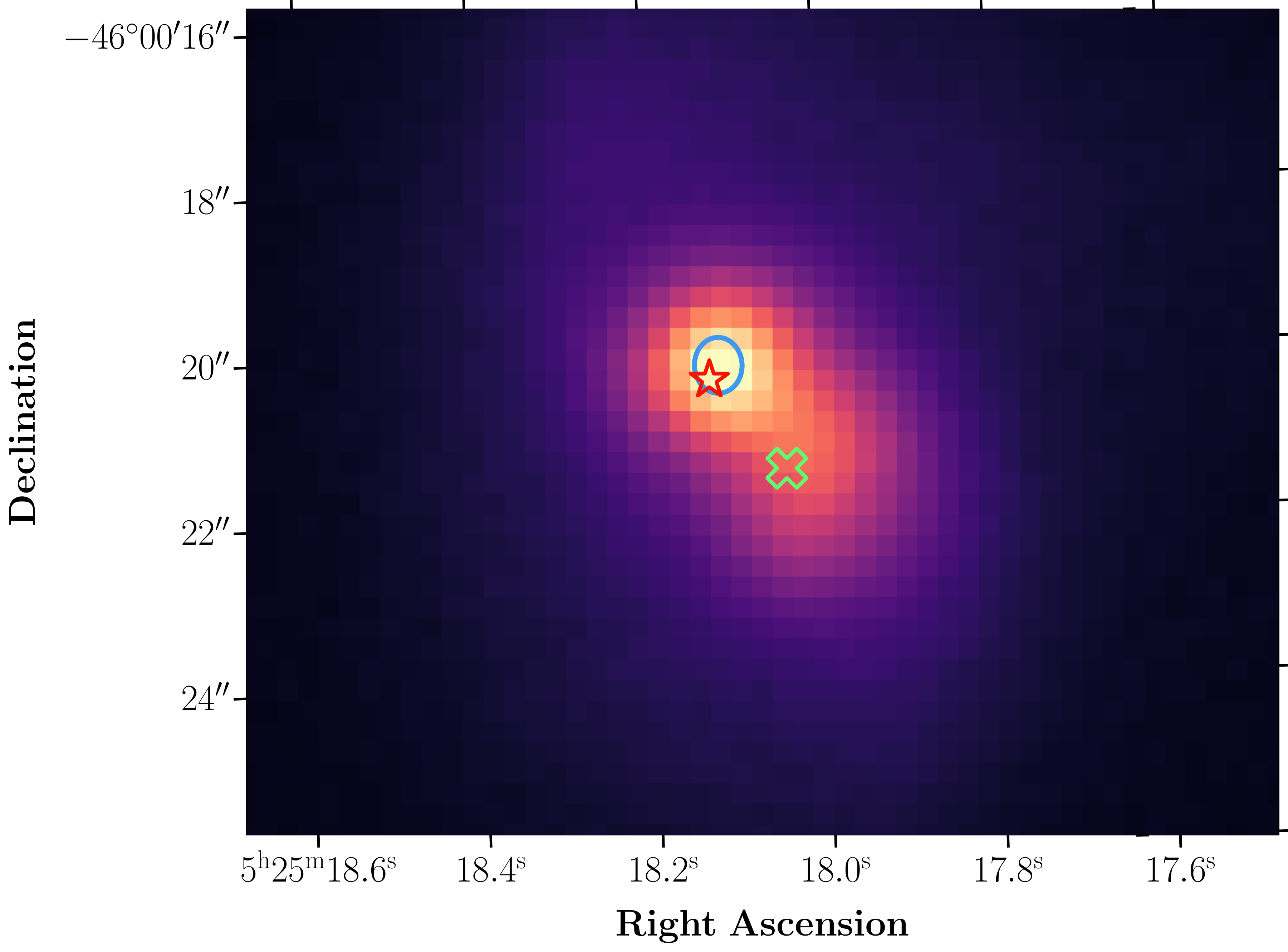}
    \caption{Pan-STARRS1 $i$-band reference image of \gal{} with a blue ellipse showing the average position of ASASSN-14ko's flares in the Pan-STARRS1 difference images. The widths of the ellipse correspond to the coordinate standard deviations. The red star denotes the position originally reported for ASASSN-14ko by \citet{holoien14ATELc}, and the green cross shows the position of the southwestern nucleus \citep{tucker2020}. The brighter, northeastern nucleus is the source of ASASSN-14ko's periodic flares.  }
    \label{fig:panstarrs}
\end{figure}

Figure \ref{fig:panstarrs} shows the Pan-STARRS1 reference image of the nuclear region of \gal{}. The ASAS-SN position for ASASSN-14ko \citep{holoien14ATELc} is marked and consistent with the northern nucleus. The Pan-STARRS1 data also shows that the location of the flares corresponds to the northern nucleus. The average and standard deviation of the Pan-STARRS1 difference image source positions is shown by the ellipse in Figure \ref{fig:panstarrs}. Its coordinates in right ascension and declination are $81.3255 \pm 0.0002 ^{\circ}$ and $-46.0056 \pm 0.0002 ^{\circ}$.

\section{X-ray, UV, and Optical Evolution} \label{xrayuvotevolution}

The September 2020 flare was the first opportunity to directly compare the X-ray and UV evolution between events. In addition, this was the first outburst where the UV rise was observed. These observations allow us to characterize the high energy properties of ASASSN-14ko's flares in a wider context, instead of as a singular episode. 

\subsection{UV/Optical Evolution} \label{uvopticalevol}

\begin{figure*}
    \centering
    \includegraphics[width=0.87\linewidth]{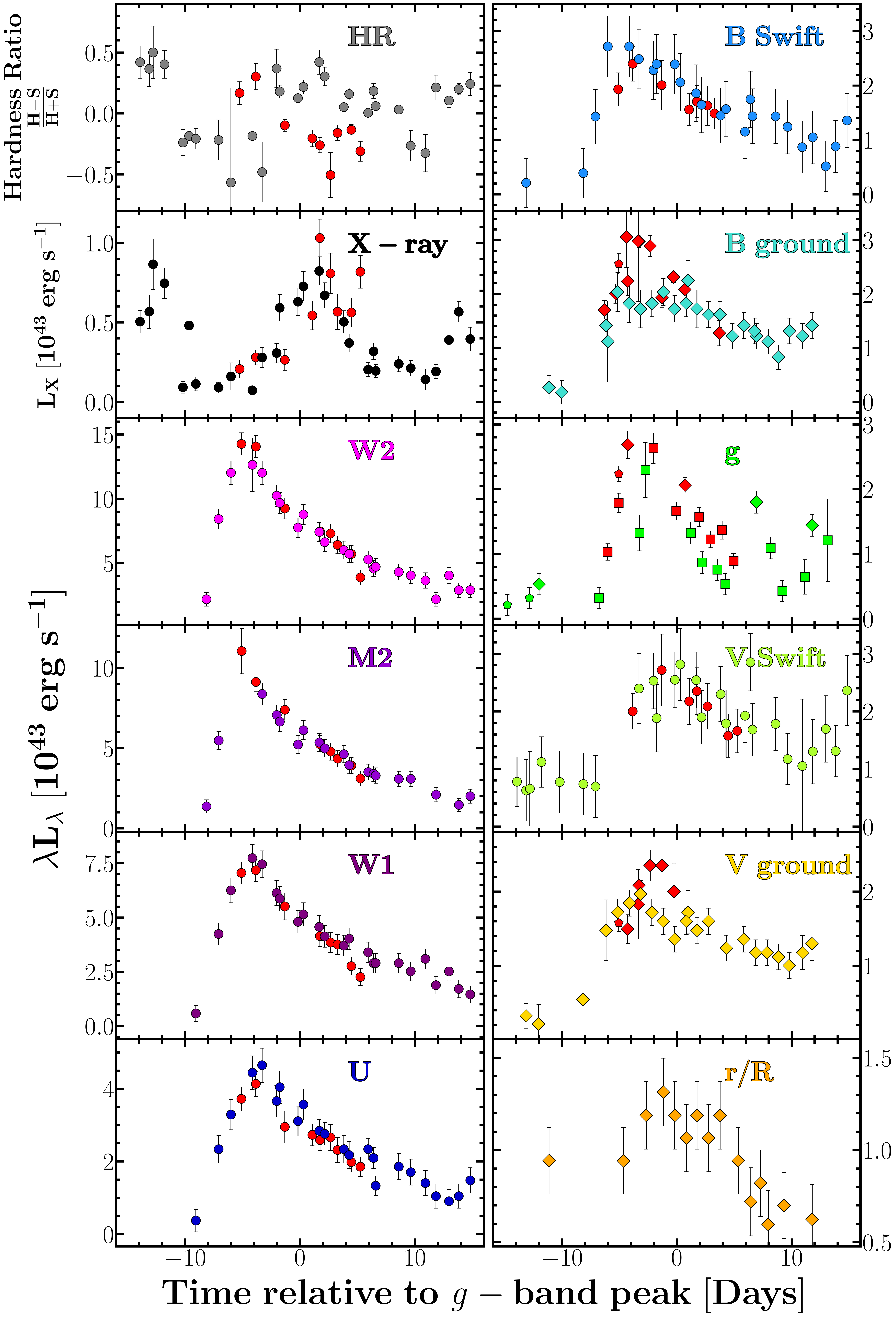}
    \caption{Host-subtracted UV/optical and 0.3-10.0 keV integrated X-ray photometry of ASASSN-14ko. The September 2020 flare data is shown in colors other than red, and the May 2020 flare data is shown in red. The time is relative to the predicted times for the peaks in the updated timing model of Section \ref{uvopticalevol}. The peak time for the September data is MJD 59098.88, and the peak time for the May data is MJD 58988.75. Circles indicate \textit{Swift} data, diamonds indicate amateur astronomer data, squares indicate ASAS-SN data, and pentagons indicate LCOGT data. }
    \label{fig:hostsub_phasefold_LC}
\end{figure*}

Figure \ref{fig:hostsub_phasefold_LC} shows the host-subtracted data from September 2020 and also given in Table \ref{tab:photometry}. The host magnitudes were determined from quiescent May 2020 data derived in \citet{payne2020}. We used the September flare's peak timing to update the parameters in the timing model. Following the same procedure as \citet{payne2020}, we fit the ASAS-SN $g$-band light curve with a fifth-order polynomial and found the peak time of MJD $59097.76^{+0.88}_{-1.38}$. The errors on the peak time were measured by bootstrap resampling the light curve. Combining this time with the previous seventeen reported in \citet{payne2020} and fitting them using the peak timing equation from \citet{payne2020} defined as

\begin{equation}
    t = t_0 + nP_0 + \frac{1}{2}n^2 P_0 \dot{P} + \frac{1}{6} n^3 P_0 \dot{P}^2,
\end{equation}
gives reference time $t_0 = 4.9 \pm 2.6$, mean period $P_0 = 114.4 \pm 0.4$ days, and a period derivative $\dot{P} = -0.002 \pm 0.0003$. The errors on the peak times were expanded in quadrature to get a $\chi^2$ per degree of freedom of unity. These updated parameters predict the next flare to peak in the $g$-band on MJD $59208.76 \pm 1.1$.

\begin{figure}
    \centering
    \includegraphics[width=1.02\linewidth]{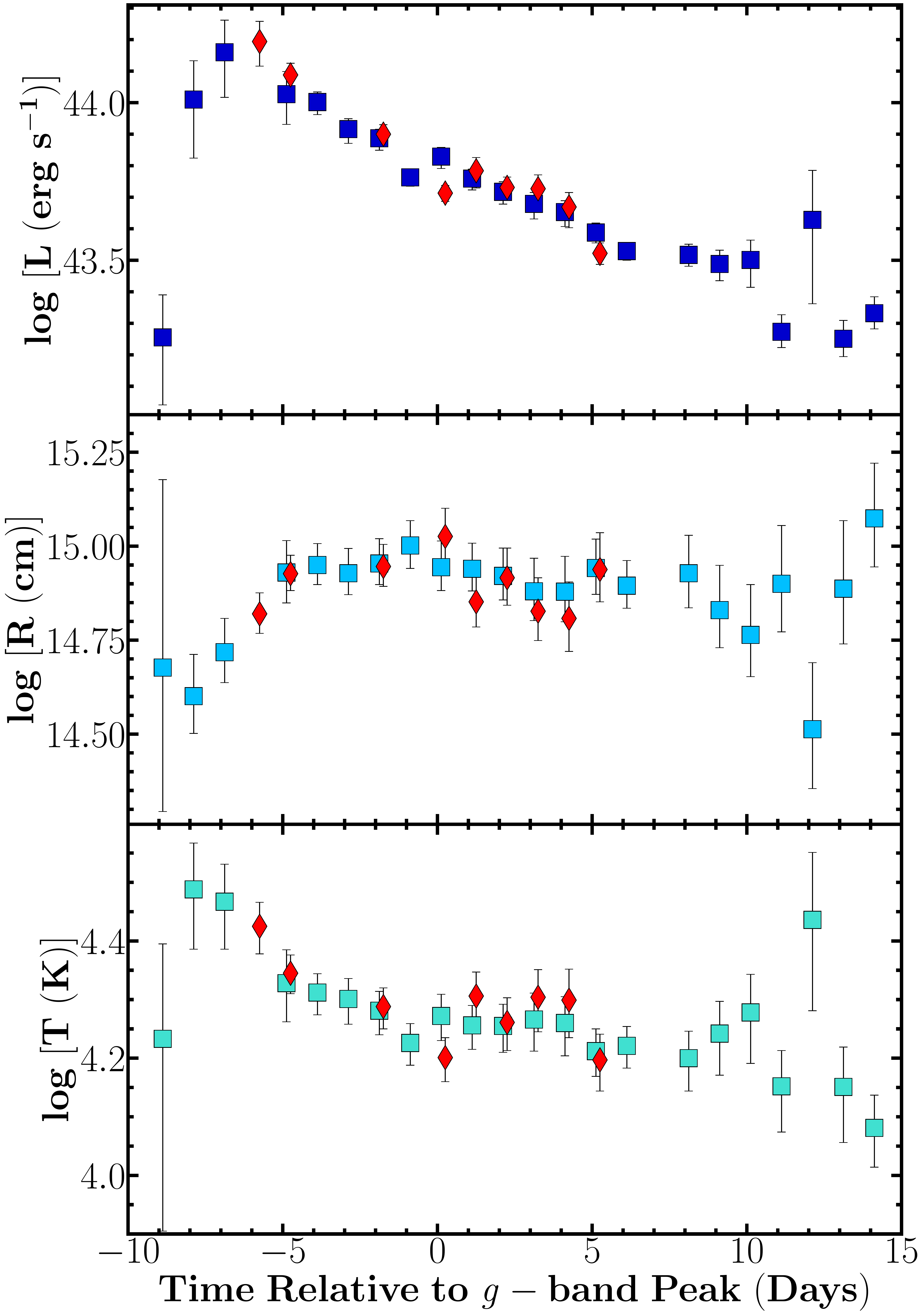}
    \caption{Evolution of ASASSN-14ko's luminosity (\textit{top panel}), effective radius (\textit{middle panel}), and temperature (\textit{bottom panel}) from blackbody fits to the host-subtracted UV/optical \textit{Swift} data. The September 2020 flare data are shown by squares in shades of blue and the May 2020 flare data are shown by red diamonds.  }
    \label{fig:bbody_evol}
\end{figure}

Figure \ref{fig:hostsub_phasefold_LC} also shows the photometry from the May 2020 flare, where we have aligned them using the predicted peak times from the updated timing model. The similarities of the flares across the UV/optical is remarkable. We then fit each epoch using a blackbody model to track the change in blackbody luminosity, temperature, and radius for the duration of the flare. As shown in Figure \ref{fig:bbody_evol}, the blackbody temperature and luminosity evolved rapidly during the rise, peaking within two days. Considering this rapid early evolution, sub-daily or 6-hour cadence will be a crucial requirement for future observing campaigns to provide better constraints on any potential  short-timescale changes and to look for differences between the peak times of different wavelengths. 

\begin{figure*}
    \centering
    \includegraphics[width=\linewidth]{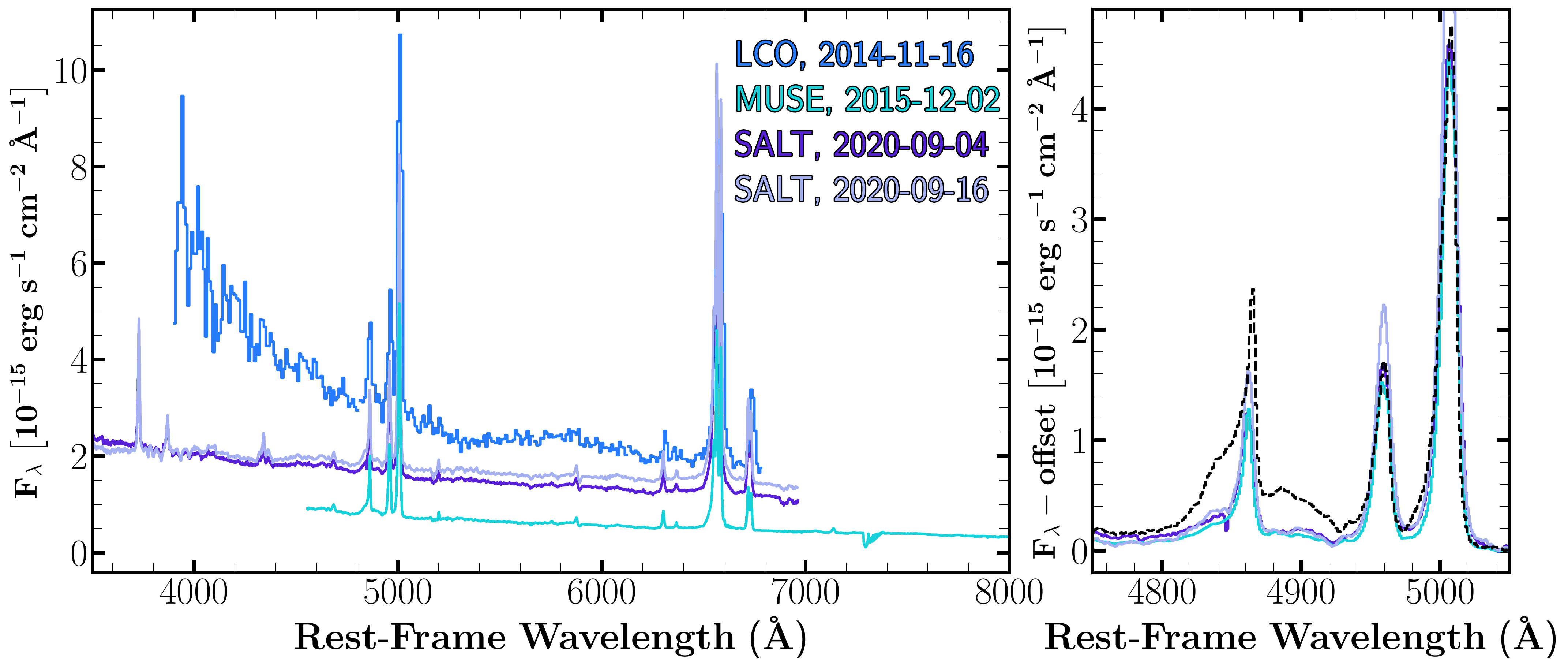}
    \caption{(\textit{left}) Comparison of the SALT ASASSN-14ko spectra observed during the September 2020 flare (dark purple and light purple), an archival spectrum observed during quiescence obtained with MUSE (turquoise), and the spectrum of the discovery flare in 2014 (blue) \citep{holoien14ATELc}. (\textit{right}) The continuum-subtracted SALT and MUSE spectra centered around H-$\beta$ and [OIII] compared with the MUSE spectrum extracted from the northeastern nuclei of \gal{} only (black, dashed), as shown in \citet{tucker2020}.  }
    \label{fig:2020septspectra}
\end{figure*}

Figure \ref{fig:2020septspectra} compares the SALT spectra and the MUSE spectrum obtained during quiescence in 2015 that was analyzed in \citet{tucker2020}. We extracted a spectrum from the MUSE data cube using a rectangular region of width $1.\!\!''5$ matching the geometry of the SALT slit on UT 2020-09-04. The blue continuum is similar but shallower than what was observed for the original discovery spectrum taken 7 days post peak \citep{holoien14ATELc}. There was also a small degree of broadening in H$\beta$ during the flare. However, we caution that the nuclear region is especially complex \citep{tucker2020} and the extracted spectra depend on seeing, slit width, slit angle, and extraction region as illustrated by the right panel of Figure \ref{fig:2020septspectra}.  Future work to robustly measuring changes in the line profiles during an outburst will require integral field unit spectroscopy or careful consideration of the slit placement.

\begin{table}[t]
    \centering
    \begin{tabular*}{\columnwidth}{l l l }
\toprule

\hline

JD    &    Band    &    $L_X$ or $\lambda L_{\lambda}$ [10$^{43}$ erg s$^{-1}$] \\
\hline
2459089.823    &    X-ray    &    0.114   $\pm$    0.043\\
2459090.746    &    $W2$    &    2.196    $\pm$    0.540\\
2459090.752    &    $M2$    &    1.375    $\pm$    0.413\\
2459089.823    &    $W1$    &    0.584    $\pm$   0.362\\
2459089.824    &    $U$    &    0.374    $\pm$    0.308\\
2459091.806    &    $B$    &    1.428    $\pm$    0.501\\
2458982.732    &    $g$    &    1.029    $\pm$   0.128\\
2459087.100    &    $V$    &    1.121    $\pm$   0.440\\
2459087.771    &    $R/r$    &    0.942    $\pm$   0.181\\

\hline 

\end{tabular*}
    \caption{Light curve data of ASASSN-14ko used in this analysis. Only one observation in each band is shown here to demonstrate its form and content. Table to be published in its entirety in machine-readable form in the online journal. }
    \label{tab:photometry}
\end{table}

\subsection{X-ray Light Curve Evolution}

As with the UV/optical emission, the X-ray flux evolved in a consistent manner between the May and September 2020 flares, as shown in Figure \ref{fig:hostsub_phasefold_LC}. However, the X-ray evolution contrasts sharply with the UV/optical behavior. The X-ray flux first decreased during the UV/optical rise, and then increases as the UV/optical begins to decline. The initial drop in the X-ray flux is very rapid, by a factor of $\sim$9 in $\sim$2.6 days. The second dip in this ``double-dip" behavior was similarly rapid, with a decrease in the flux by a factor of $\sim$4 in $\sim$4.3 days. Comparing the May 2020 and September 2020 events indicates that the drop in X-ray flux during the UV/optical rise is a characteristic feature of both flares. 

\begin{figure}
    \centering
    \includegraphics[width=\linewidth]{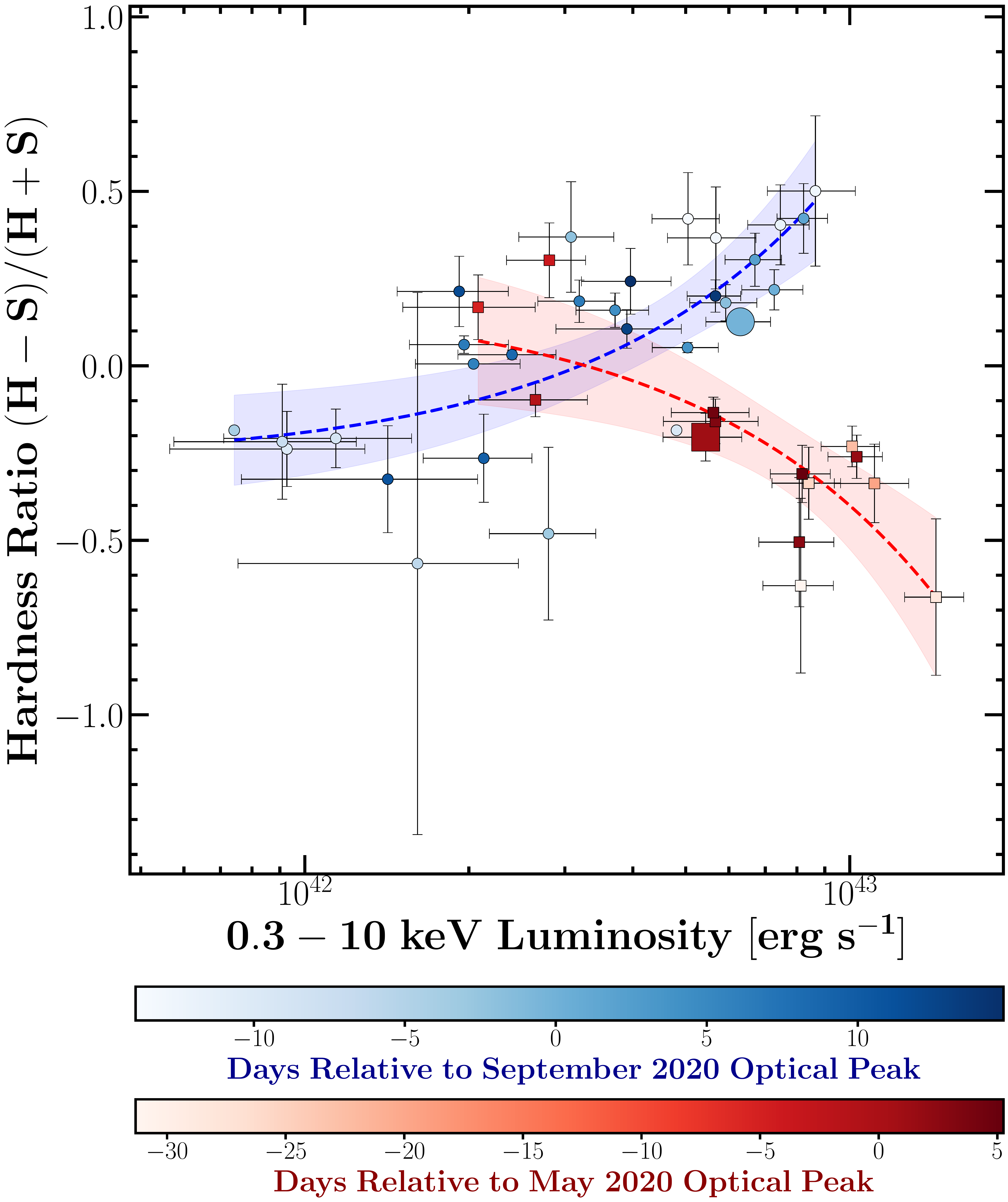}
    \caption{Hardness ratio evolution for the May 2020 (red squares) and September 2020 (blue circles) ASASSN-14ko outbursts. The two larger points denote the data closest in time to the optical peaks of each flare. The dotted lines with corresponding surrounding transparent regions show the best-fit linear regressions and 95\% confidence intervals.  Comparing the two flares reveals the two events had distinct X-ray properties in terms of the hardness ratios at peak optical flux and the hardness ratios at peak X-ray flux. The September flare was harder at peak luminosity, whereas the May flare was softer at peak luminosity. }
    \label{fig:Xray_lum_HR}
\end{figure}

While the evolution of the total X-ray flux is very similar between the two flares, the hardness ratios evolve quite differently, as shown in Figure \ref{fig:Xray_lum_HR}. In May 2020, the peak X-ray luminosities corresponded to the lowest hardness ratios implying relatively softer X-ray emission when the source is X-ray brightest, whereas the peak luminosities in September 2020 corresponded to the highest hardness ratios implying much harder X-ray emission when the source is X-ray brightest. In addition, the X-ray flux minimum during UV/optical peak coincided with an increased hardness ratio in May 2020 whereas it coincided with a decreased hardness ratio in September 2020. The hardness ratio at the optical peak in May 2020 was $-0.20 \pm 0.07$, whereas the hardness ratio at the optical peak in September 2020 was $0.13 \pm 0.03$, as shown by the larger points in Figure \ref{fig:Xray_lum_HR}. 
 
\subsection{X-ray Spectra} 

\begin{figure*}
    \centering
    \includegraphics[width=\linewidth]{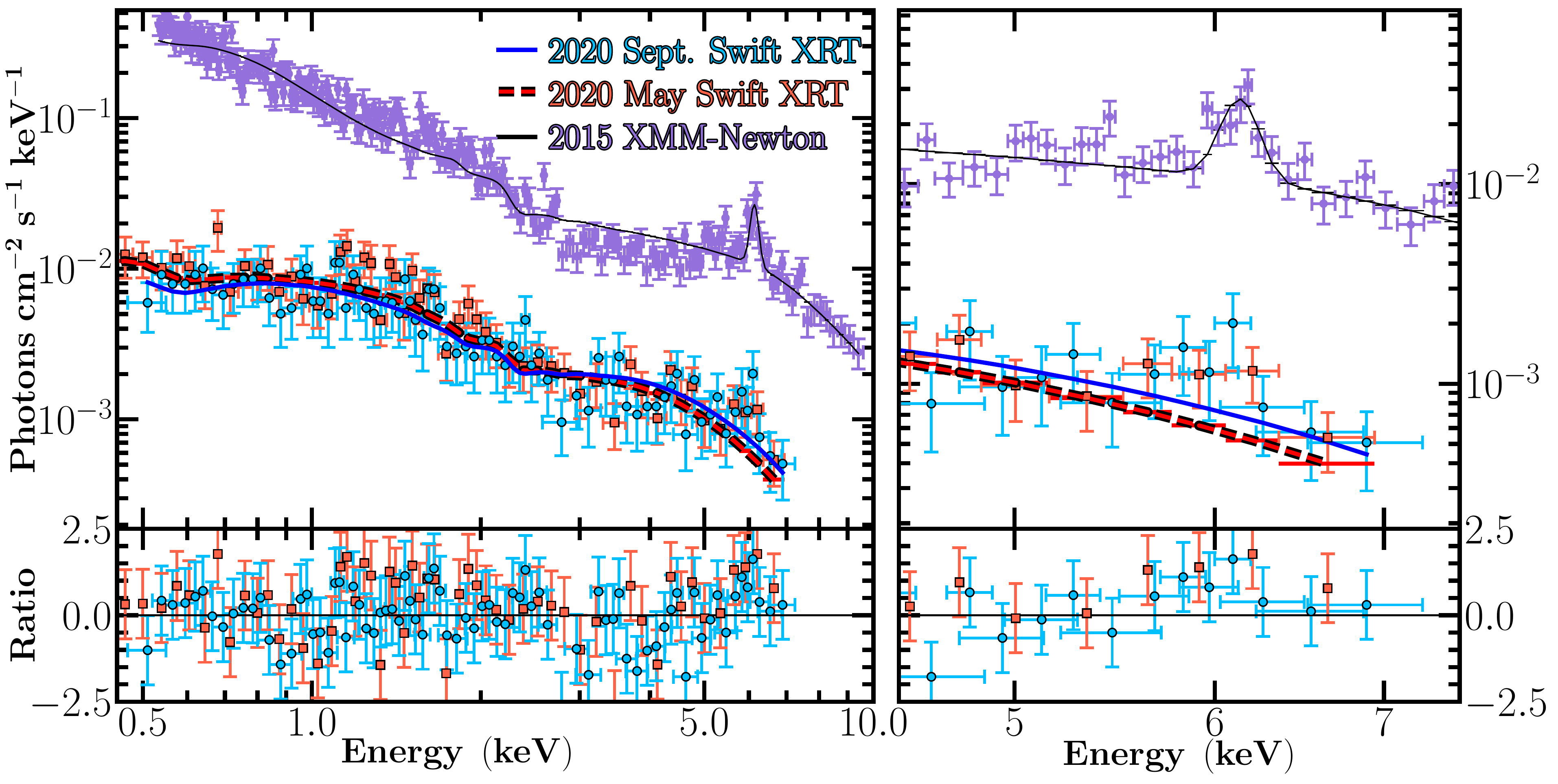}
    \caption{Stacked \textit{Swift} X-ray spectra for the September (blue circles) and May  (red squares) flares compared to archival \textit{XMM-Newton} spectrum in purple that was discussed in \citet{payne2020}. The combined 2020 data are best fit by a power law plus blackbody with photon index of $0.68 \pm 0.08$ and a  temperature of $0.18 \pm 0.02$ keV with a reduced $\chi^2$ of 1.22. The September spectrum is very similar to the spectrum from May, which was best-fit with a single power law with photon index $1.09 ^{+0.20} _{-0.22}$ but the fit was not improved by adding an additional blackbody with temperature $0.13 \pm 0.03$ keV. Shown in the right panel are the same spectra but zoomed into the region surrounding the Fe K $\alpha$ line near 6.4 keV. }
    \label{fig:xrayspectra}
\end{figure*}

Given the numbers of counts, we can only study the average X-ray spectrum in detail. Figure \ref{fig:xrayspectra} compares the stacked \textit{Swift} spectra of the May 2020 and September 2020 flares. The best-fit to the September 2020 X-ray data was a power law plus blackbody with photon index of $0.68 \pm 0.08$  and temperature of $0.18 \pm 0.02$ keV with a reduced $\chi^2$ of 1.20. The unabsorbed fluxes in the 0.5-7.0 keV energy band for each of these components is $(7.5\pm0.2)\times10^{-13}$ erg cm$^{-2}$ s$^{-1}$ and $(1.1\pm0.6)\times10^{-13}$ erg cm$^{-2}$ s$^{-1}$, respectively. For fitting the spectra, the absorbing column density was set to $3.5 \times 10^{20}$ cm$^{-2}$, the Galactic column density along the line of sight. However, the data were equally well fit by two power laws with photon indices of $2.17 \pm 0.12$ and $0.10 \pm 0.15$ with a reduced $\chi^2$ of 1.20. Assuming a single power law was not a very good fit, with a reduced $\chi^2$ of 1.70. 

As reported in \citet{payne2020}, the best-fit model to the merged \textit{Swift} data from May 2020 was a single power law with photon index $1.09 ^{+0.20} _{-0.22}$ and the fit was not significantly improved by adding an additional blackbody with temperature $0.13 \pm 0.03$ keV. However, the data from May 2020 also had fewer total counts due to the lower total integration time (22.5 ksec versus 56.0 ksec). Overall, even though the blackbody plus power law and the two power laws fit the spectrum well, a power law with a photon index of 0.10 plus blackbody is more aligned with the expectation of AGN which have blackbody temperatures of  $>0.1$ keV. 

Archival \textit{XMM-Newton} data of \gal{} taken during ASASSN-14ko's quiescence in 2015 show a prominent Fe K $\alpha$ line near 6.4 keV, as shown in Figure \ref{fig:xrayspectra}. There is also a small increase in flux at this energy level present in the \textit{Swift} XRT spectra, however there are too few counts in the \textit{Swift} observations to properly constrain the presence of this line. Due to the spatial resolution of \textit{Swift} and \textit{XMM-Newton}, we are unable to resolve the two distinct AGN nuclei in \gal{}, making it difficult to disentangle the X-ray emission between the two sources. Future observations with higher spatial resolution using \textit{Chandra} are needed to differentiate between the two nuclei, which will also indicate the physical extent of the Fe K $\alpha$ line emission in \gal{}.

We also separately considered the average X-ray spectra before and after the UV/optical peak. Before the peak, we find that an absorbed power law plus blackbody model is a better fit than two power laws or a single power law model. This model consists of a photon index of $0.91 \pm 0.1$ and a temperature of $0.16 \pm 0.03$ keV with a reduced $\chi^2$ of 1.5. The best post-UV/optical peak model is also an absorbed power law plus blackbody but with a photon index of $0.62 \pm 0.09$ and a temperature of $0.18 \pm 0.02$ keV with a reduced $\chi^2$ of 1.1. Comparing the spectra between these two epochs shows that after UV/optical peak, the photon index decreased and became harder. This is consistent with the hardness ratio evolution shown in Figure \ref{fig:hostsub_phasefold_LC}. 

\vspace{1cm} 

\section{Discussion} \label{discussion}

The outbursts of ASASSN-14ko continue to follow a periodic trend. However, the updated timing model parameters $t_0$, $P_0$, and $\dot{P}$  changed compared to those found in \citet{payne2020}, although the $g$-band optical peak occurred within error bars on the prediction in \citet{payne2020}. Compared to May 2020, $\dot{P}$ changed by $-0.0003$. Seeing a change in $\dot{P}$ between events seems to indicate that the system does not have a global $\dot{P}$ and the orbit of a partial TDE should show some stochasticity in its evolution. Regardless of this variation, the timing model  still describes the system well  and was able to  correctly predict the next flare in December 2020 (A.V. Payne et al., in prep). 

We now see in two flares that the rapid rise and fall observed in the optical over the last 6 years is also typical of the UV emission. This $\sim$20 day timescale over which the flares' blackbody luminosity decays is much shorter than for all previously studied TDEs. In the \citet{hinkle2021, hinkle20a} studies of the correlation between peak luminosity and decline rates in TDEs, all of the examples decline on timescales of over 100 days.

While the optical, UV, and X-ray luminosity evolution of the individual flares are nearly identical, the X-ray spectra of the May and September flares evolved differently. In May, the spectra softened as the X-ray luminosity increased, while in September they hardened. This  is the first property found that distinguishes individual flares.  \citet{auchettl2017} systematically studied a sample of TDEs in order to uniformly characterize the X-ray emission of TDEs as a whole. \citet{auchettl2017} found that all TDE candidates had soft X-ray emission with HRs ranging from $-1.0$ to $+0.3$ and the majority of objects had HRs between $-1.0$ and $0$ while highly variable AGN have hardness ratios ranging from $-1$ to $+1$. Thus, both flares are consistent with previously detected TDE candidates even though the HR evolutions are distinct.

ASASSN-14ko's X-ray behavior is similar to another nuclear transient discovered by the ASAS-SN survey, ASASSN-18el  \citep{nicholls2018}.  ASASSN-18el displayed strong Balmer emission lines that broadened after the galaxy's nucleus brightened and was subsequently described as a changing-look AGN \citep{trakhtenbrot201918el}. Further X-ray monitoring by \citet{ricci2020} revealed strong variability on longer timescales. The X-ray luminosity decreased rapidly $\sim$160 days after the optical event and then slowly increased to eventually exceed its pre-flare level. This led to an X-ray light curve characterized by an asymmetric trough sharply contrasting with the optical brightening that triggered the event. While the overall pattern is similar, the timing of the X-ray changes relative to the UV/optical is different in ASASSN-14ko.  ASASSN-18el's X-ray trough occurred much later after the UV/optical brightening, whereas ASASSN-14ko's X-ray flare decreased during the UV/optical rise and peak, with a secondary dip later in the UV/optical decline. In addition, ASASSN-18el occurred over hundreds of days as opposed to ASASSN-14ko's much faster evolution where the X-ray flux decrease occurred over only $\sim$14 days. 

\citet{ricci2020} also found that the X-ray flux variability of ASASSN-18el and spectral variability was tightly connected, becoming harder at times of increased X-ray flux. ASASSN-14ko showed the same trend in September 2020 but not in May 2020. ASASSN-18el's spectrum was dominated by a blackbody-like component after the optical brightening, with the previously-observed power law component faded during the decreased X-ray flux \citet{ricci2020}. When the X-ray flux increased again, the power law component returned. \citet{ricci2020} argued that the disappearance and reappearance of the power law component was evidence that the X-ray corona was destroyed by a TDE and then gradually reformed. This is another facet that contrasts with ASASSN-14ko, whose X-ray spectrum was consistently modeled as a power law plus blackbody for the duration of the September 2020 event. That ASASSN-14ko's power law component remains present throughout the event may suggest that the innermost regions of the accretion flow  remains intact as opposed to being destroyed by the TDE.

Changes in spectral lines may be clear discriminators between TDE or AGN activity, or key features defining new classes of objects (e.g., \citealt{trakhtenbrot2019nature}, \citealt{frederick2020}, \citealt{neustadt2020}, \citealt{vanvelzen2021}). Reverberation mapping (\citealt{blandford82}, \citealt{peterson93}, \citealt{peterson04}) is a technique that has been used for decades as a powerful method to probe the inner-most regions of AGN. The timings of ASASSN-14ko's future flares can be predicted, allowing high-cadence spectroscopic campaigns to be organized around future events that would be able to monitor changes in the spectral lines over the flare and how they correlate with changes in the continuum. AGN broad lines typically become narrower as they become more luminous due to the lower Keplerian velocity at greater distances. TDEs, however, show the opposite behavior. Previously studied TDEs have shown that optical emission lines strengthen and broaden with increased luminosity. For example, both a blue continuum and prominent, broad emission lines HeII $\lambda4686$, H$\beta$, and H$\alpha$ formed in the TDEs ASASSN-14ae \citep{holoien14b} and ASASSN-14li (\citealt{jose2014}, \citealt{holoien16}). For ASASSN-14ae they narrowed and then faded completely after $\sim$750 days \citep{brown2016} while they were still prominent at $\sim$500 days for ASASSN-14li \citep{brown17}. This behavior was also apparent in ``Helium-rich" TDE ASASSN-15oi (\citealt{holoien16b}, \citealt{holoien2018}), and Bowen TDE ASASSN-18pg (\citealt{leloudas2019}, \citealt{holoien2020}). 

Future spectroscopic campaigns need to carefully consider the complex host galaxy background at the center of \gal{}.  A spectroscopic observing campaign using long-slit spectrographs would require careful consideration of the slit orientations to either avoid contamination from the southern nucleus or try to always have the same level of contamination. Ideally, the observations would use an IFU spectrograph where the extracted aperture and background can be selected. 

\vspace{1cm}

\section{Conclusions} 
\label{conclusions}

ASASSN-14ko continues to be a periodic nuclear transient at the center of AGN \gal{} whose flares can be predicted with a timing model consisting of a mean period and a non-zero, negative period derivative. The September 2020 flare event was the first opportunity to directly compare individual outbursts at X-ray and UV wavelengths. In this work, we show that: 

\begin{itemize}
    \item The brighter, northeastern nucleus of \gal{} is the source of the flares. 
    \item ASASSN-14ko's updated timing model parameters are $t_0 = 4.9 \pm 2.6$, $P_0 = 114.4 \pm 0.4$ days, and $\dot{P} = -0.002 \pm 0.0003$ after including the latest peak timing from September 2020, which occurred on MJD $59097.76^{+0.88}_{-1.38}$. The model predicted a next peak for December 2020 on MJD $59208.76 \pm 1.1 $ which occurred on MJD $59205.65 \pm 0.19$ (A.V. Payne et al., in prep). 
    \item When comparing the May 2020 and September 2020 X-ray and UV light curves, the two flares show a rapid UV/optical brightening coincident with an equally rapid decrease in X-ray flux.  
    \item The X-ray hardness ratio evolution is currently the only clear difference between the May and September flares. The September flare was harder at peak X-ray luminosity, whereas the May flare was softer at peak X-ray luminosity.
    
\end{itemize}

Future flares will continue to be important to determine which characteristics change and what remain the same, especially in the high energy regime. One aspect in particular is determining if there will be any continued difference in the X-ray hardness ratio and spectral properties. In addition, future reverberation mapping campaigns centered around the predicted flare timings could provide an accurate black hole mass measurement, especially when using IFU instruments to avoid contamination from the southwestern nucleus in the host galaxy.  

{\bf Software:}
astropy \citep{astropy2018}, ftools \citep{blackburn95}, HEAsoft \citep{heasarc2014}, IRAF \citep{tody1986, tody1993}, numpy \citep{harris2020}, matplotlib \citep{hunter2007} 

\acknowledgments
We thank Jennifer van Saders for helpful discussion. 
We thank the Las Cumbres Observatory and its staff for its continuing support of the ASAS-SN project. ASAS-SN is supported by the Gordon and Betty Moore Foundation through grant GBMF5490 to the Ohio State University, and NSF grants AST-1515927 and AST-1908570. Development of ASAS-SN has been supported by NSF grant AST-0908816, the Mt. Cuba Astronomical Foundation, the Center for Cosmology and AstroParticle Physics at the Ohio State University, the Chinese Academy of Sciences South America Center for Astronomy (CASSACA), and the Villum Foundation. 

A.V.P. acknowledges support from the NASA Graduate Fellowship through grant 80NSSC19K1679. B.J.S., C.S.K., and K.Z.S. are supported by NSF grant AST-1907570. B.J.S. is also supported by NASA grant 80NSSC19K1717 and NSF grants AST-1920392 and AST-1911074. C.S.K. and K.Z.S. are supported by NSF grant AST-181440. J.T.H. is supported by NASA award 80NSSC21K0136. Support for T.W.-S.H. was provided by NASA through the NASA Hubble Fellowship grant HST-HF2-51458.001-A awarded by the Space Telescope Science Institute, which is operated by the Association of Universities for Research in Astronomy, Inc., for NASA, under contract NAS5-26555. Parts of this research were supported by the Australian Research Council Centre of Excellence for All Sky Astrophysics in 3 Dimensions (ASTRO 3D), through project number CE170100013. T.A.T. is supported in part by Scialog Scholar grant 24215 from the Research Corporation. M.A.T. acknowledges support from the DOE CSGF through grant DE-SC0019323.  The SALT observations presented here were made through Rutgers University program 2020-1-MLT-007 and supported by NSF grant AST-1615455 to S.W.J.

Operation of the Pan-STARRS telescopes is primarily supported by the National Aeronautics and Space Administration under Grant No. NNX12AR65G and Grant No. NNX14AM74G issued through the SSO Near-Earth Object Observations Program. The PanSTARRS1 Surveys (PS1) and the PS1 public science archive have been made possible through contributions by the Institute for Astronomy, the University of Hawaii, the Pan-STARRS Project Office, the Max-Planck Society and its participating institutes, the Max Planck Institute for Astronomy, Heidelberg and the Max Planck Institute for Extraterrestrial Physics, Garching, The Johns Hopkins University, Durham University, the University of Edinburgh, the Queen’s University Belfast, the Harvard-Smithsonian Center for Astrophysics, the Las Cumbres Observatory Global Telescope Network Incorporated, the National Central University of Taiwan, the Space Telescope Science Institute, the National Aeronautics and Space Administration under Grant No. NNX08AR22G issued through the Planetary Science Division of the NASA Science Mission Directorate, the National Science Foundation Grant No. AST–1238877, the University of Maryland, Eotvos Lorand University (ELTE), the Los Alamos National Laboratory, and the Gordon and Betty Moore Foundation.

\bibliographystyle{aasjournal}
\bibliography{references}

\end{document}